\def\prd{Phys. Rev. D}

\def\apj{Astrophys. J.}
\def\apjl{Astrophys. J. Lett.}
\def\apjs{Astrophys. J.Suppl.}
\def\mnras{Mon. Not. R. Astr. Soc.}
\def\aap{Astr. Astrophys.}

\def\jcap{JCAP}

\def\procspie{Proceedings of SPIE}
\def \<{\langle}
\def \>{\rangle}
\def\approxprop{
  \def\p{
    \setbox0=\vbox{\hbox{$\propto$}}
    \ht0=0.6ex \box0 }
  \def\s{
    \vbox{\hbox{$\sim$}}
  }
  \mathrel{\raisebox{0.7ex}{
      \mbox{$\underset{\s}{\p}$}
    }}
}

\newcommand{\ra}{\;\raise1.0pt\hbox{$'$}\hskip-6pt\partial\;}
\newcommand{\lo}{\;\overline{\raise1.0pt\hbox{$'$}\hskip-6pt\partial}\;}

\newcommand{\degree}{^\circ}

\newcommand{\Abs}[1]{\begin{abstract} #1 \end{abstract}}
\newcommand{\Ack}[1]{\begin{acknowledgments} #1 \end{acknowledgments}}
\newcommand{\mktt}{\maketitle}
\documentclass[twocolumn,amsmath,amssymb,floatfix,superscriptaddress,showkeys,prd,nofootinbib]{revtex4}
\usepackage{graphicx,epsfig,natbib,color,times,bm,amsmath,multirow,hyperref,harpoon,snapshot}
\usepackage[ddmmyyyy,hhmmss]{datetime}
\hypersetup{linktocpage}

\begin{document}

\title{Methods for pixel domain correction of EB leakage}

\author{Hao Liu}\email[]{liuhao@nbi.dk}
\affiliation{The Niels Bohr Institute and Discovery Center, Blegdamsvej 17,
DK-2100 Copenhagen, Denmark}
\affiliation{Key laboratory of Particle and Astrophysics, Institute of High
Energy Physics, CAS, 19B YuQuan Road, Beijing, China, 100049}

\author{James Creswell}\email[]{james.creswell@nbi.ku.dk}
\affiliation{The Niels Bohr Institute \& Discovery Center, Blegdamsvej 17,
DK-2100 Copenhagen, Denmark}

\author{Sebastian {von Hausegger}}\email[]{s.vonhausegger@nbi.dk}
\affiliation{The Niels Bohr Institute \& Discovery Center, Blegdamsvej 17,
DK-2100 Copenhagen, Denmark}

\author{Pavel Naselsky}\email[]{naselsky@nbi.dk}
\affiliation{The Niels Bohr Institute \& Discovery Center, Blegdamsvej 17,
DK-2100 Copenhagen, Denmark}

\Abs{

In observation of the cosmic microwave background (CMB) polarization,
``$EB$~leakage'' refers to the artificial $B$-mode signal coming from the
leakage of $E$-mode signal when part of the sky is unavailable or excluded.
Correction of such leakage is one of the preconditions for detecting
primordial gravitational waves via the CMB $B$-mode signal. In this work, we
design two independent methods for correcting the $EB$~leakage directly in the
pixel domain using standard definitions of the $E$- and $B$-modes. The two
methods give consistent results, and both are fast and easy to implement.
Tests on a CMB simulation containing zero initial $B$-mode show an efficient
suppression of the $EB$ leakage. When combined with the MASTER method to
reconstruct the full-sky $B$-mode spectrum in simulations with a relatively
simple mask, the error from EB-leakage is suppressed further by more than one
order of magnitude at the recombination bump, and up to three orders of
magnitude at higher multipoles, compared to a ``pure'' MASTER scheme under the
same conditions. Meanwhile, although the final power spectrum estimation
benefits from apodization, the pixel domain correction itself is done without
apodization, and thus the methods offer more freedom in choosing an
apodization based on specific requirements.

}


\mktt

\section{Introduction}\label{sec:intro}

The analysis of the Cosmic Microwave Background (CMB) polarization lies in the
focus of current CMB missions such as Planck, as well as future
missions~\citep{2014A&A...571A...1P, 2016A&A...594A...1P, 2018arXiv180706205P,
2012SPIE.8442E..19H, 2016arXiv161002743A, 2011arXiv1110.2101K, quijote2012,
doi.10.1093.nsr.nwy019}. The main scientific goal is to detect primordial
gravitational waves in the $B$-mode of the polarized CMB signal. The
contribution of primordial gravitational waves is quantified by the
tensor-to-scalar ratio $r$. Currently, this parameter is constrained to
$r\lesssim0.07$~\citep{2015PhRvL.114j1301B, 2018arXiv180706209P}, and future
missions are aiming for a sensitivity of
$r\sim10^{-4}$~\citep{2016arXiv161002743A}. However, CMB missions of the near
future are all ground-based, which means that they provide only partial sky
coverage. On an incomplete sky map, the separation of the polarized signal
into $E$- and $B$-modes will be affected by ``leakage'' (the so-called
$EB$~leakage), and the resulting $B$~map can be strongly
contaminated~\cite{2002PhRvD..65b3505L, Bunn:2002df}. This kind of leakage
must be carefully corrected to reach the above target.

One way to study the $EB$~leakage due to incomplete sky coverage is by
constructing localized estimators that are associated with the mask or window
function defined on the sky fraction in question, as was first proposed
by~\cite{Bunn:2002df}, and subsequently used in many such studies,
e.g.~\cite{2003PhRvD..68h3509L, 2003NewAR..47..987B, 2004mmu..symp..309Z,
PhysRevD.82.023001, 2006PhRvD..74h3002S, 2007PhRvD..76d3001S}. Commonly such
estimators are referred to as ``pure'' $E$/$B$-modes. However, note that so
far it was not noticed how to perform a pixel domain conversion from the
``pure'' $E$- and $B$-modes to standard $E$- and $B$-modes that are defined on
the full sky. It is important to note that only such full-sky $E$- and
$B$-modes are always orthogonal to each other.

We here introduce two methods for the correction of the $EB$~leakage in the
pixel domain, which only use the standard full-sky definitions of $E$- and
$B$-modes~\cite{PhysRevD.55.1830, 1997PhRvD..55.7368K, 0004-637X-503-1-1}. The
first method is motivated by studying properties of the leakage at the mask's
boundary: in a series of works~\citep{2003NewAR..47..987B,
2011PhRvD..83h3003B, 2012MNRAS.424.1694B, 2017PhRvD..96d3523B}, the technical
details of the $EB$~leakage and their possible solutions were thoroughly
discussed, and even the idea of correcting the leakage using relaxation
methods was mentioned. In our paper, we implement the relaxation method using
diffusive inpainting, e.g.~\citep{2014A&A...571A..24P, 2016A&A...594A..17P}.
However, we also note the limitation of the relaxation method: it only gives a
particular solution that could ignore small scale features. Improvement of
this solution requires more knowledge about the real $B$-mode signal inside
the region, which is unfeasible in this way.

To this effect, we introduce an alternative and novel $EB$~leakage correction
method in the pixel domain. The $E$ and $B$~signals are regarded as being
composed of contributions from different regions of the sky. When part of the
sky is unavailable the leakage correction is carried out by recycling the
$E$-family component of the $Q$ and $U$ Stokes
parameters~\cite{2018JCAP...05..059L} derived from only the available sky
region. The two methods give similar results. We show that generally the
second method performs better, and we will therefore focus our attention on
it. However, in special cases the first method can outdo the latter as we
discuss below.

We emphasize that neither of the two methods requires prior knowledge of the
underlying $EE$ or $BB$ power spectra, which is of significant advantage. It
is also important to note that, compared to previous methods,
e.g.~\cite{2006PhRvD..74h3002S, 2007PhRvD..76d3001S, 2010A&A...519A.104K,
PhysRevD.82.023001}, this work provides the first correction of the
$EB$~leakage in the pixel domain using the standard full-sky definition of the
$E$- and $B$-modes.  We shall discuss a list of benefits in the main body of
this paper.

This paper is organized as follows: in Sec.~\ref{sec:the methods}, we
introduce the two methods and provide examples using two different masks for
illustration. Their performance is tested in Sec.~\ref{sec: evaluation of
the residual}, and a brief discussion is given in
Sec.~\ref{sec:disuss}.

\section{Methods and examples}\label{sec:the methods}

We here introduce two independent methods; the first is a relaxation method,
and the second provides a more elaborate solution utilizing our recently
introduced $EB$-families. In~\cite{2019arXiv190400451L}, we provide
mathematical proof that method 2 is the best blind correction of the
EB-leakage in the pixel-domain, but in this work we still present the results
of both methods as a useful cross-check.

For the convenience of reading, the background and principles of the two
methods will be introduced in Appendix~\ref{app:intro 1 2}, and below we
present the procedures of the two methods directly.

\subsection{Method 1: Diffusive inpainting}\label{sub:method 1}

Method 1 is to estimate the $EB$-leakage by diffusive
inpainting~\citep{2014A&A...571A..24P, 2016A&A...594A..17P}, in which sky
pixels\footnote{For this we have used the \textsc{HEALPix} package
(\url{http://healpix.sourceforge.net}) and therefore adopted their
pixelization scheme. However, the method is not tailored to function with that
pixelization only.} are iteratively replaced by the average of their
neighbors, except for the pixels on the boundary. The procedure is:
\begin{enumerate}
  \item Begin with the corrupted $B$~map derived from a masked sky (for
  calculation of the $B$~map, see Appendix~\ref{app:eb maps}).
  \item \label{itm:method 1 step 2}Set all pixels on the sky to zero except
   those at the edge of the valid region, which constitute the boundary
   condition.
  \item Perform diffusive inpainting on the valid sky as mentioned above using
  the boundary condition in step~\ref{itm:method 1 step 2}. On convergence,
  the result is a template for the $EB$~leakage.
  \item Subtract the derived template from the corrupted $B$~map in order to
  arrive at the corrected $B$~map.
\end{enumerate}

\subsection{Method 2: Recycling the E-mode}\label{sub:method 2}

Method 2 is to estimate the $EB$-leakage by recycling the $E$-mode signal. As
mentioned above, this is the best blind estimate of the $EB$-leakage in pixel
domain. The procedure is:
\begin{enumerate}
  \item \label{itm:corr step1} Begin with a sky map
  $\mathbf{P}=(Q,U)$ and a mask, calculate $\mathbf{P_E'}=(Q_E,U_E)'$ and
  $\mathbf{P_B'}=(Q_B,U_B)'$ directly from \textbf{masked} $\mathbf{P}$.
  \item \label{itm:corr step2} Similarly, obtain $\mathbf{P_B''}=(Q_B,U_B)''$
  from \textbf{masked} $\mathbf{P_E'}$.
  \item Using the same mask, $\mathbf{P_B''}$ is the template for the
  $EB$~leakage in the available region. Use it to remove the $EB$~leakage from
  $\mathbf{P_B'}$ by linear fitting.
\end{enumerate}

Above we only described the method and the procedure in terms of the $E$- and
$B$-families. Note that while it is necessary to do step~\ref{itm:corr step1}
via the $E$- and $B$-family decomposition, starting from step~\ref{itm:corr
step2}, one is also free to proceed in terms of the actual $B$-modes, and
arrive at a $B$~map as a template; both give similar results. However, for
power spectrum estimation, the $B$~map template gives slightly better results
(about $10\%$ lower error). For correcting the morphology of the corrupted map
in the pixel domain, the variant with $(Q_B,U_B)''$ is slightly better. In
this work, since we will eventually compute power spectra, all pixel domain
results will be presented in the form of $B$~maps.

\subsection{Examples and comparison}\label{sub:examples}

\begin{figure*}[ht!]
    \centering
    \includegraphics{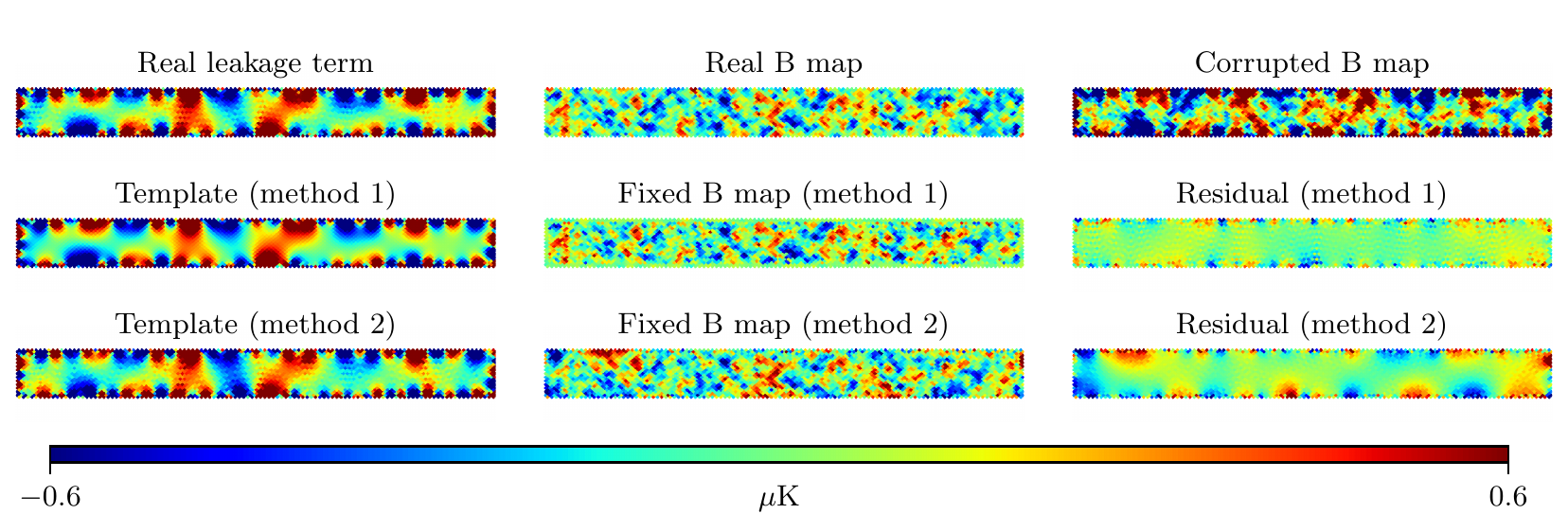}
    \caption{Examples of pixel domain $EB$~leakage corrections with a $r=0.05$
    simulation in a belt region that is $20\degree$ wide and $2\degree$ high.
    \textit{Upper panels:} The real leakage term (left), the real $B$~map
    (middle), and the corrupted $B$~map (right). \textit{Middle panels:}  The
    results of method~1.  The derived template (left), the corrected $B$~map
    (middle), and the residual leakage (right). \textit{Bottom panels:} Same
    as middle panels but for method~2.}
    \label{fig:show results}
\end{figure*}

\begin{figure}[!htbp]
    \centering
    \includegraphics{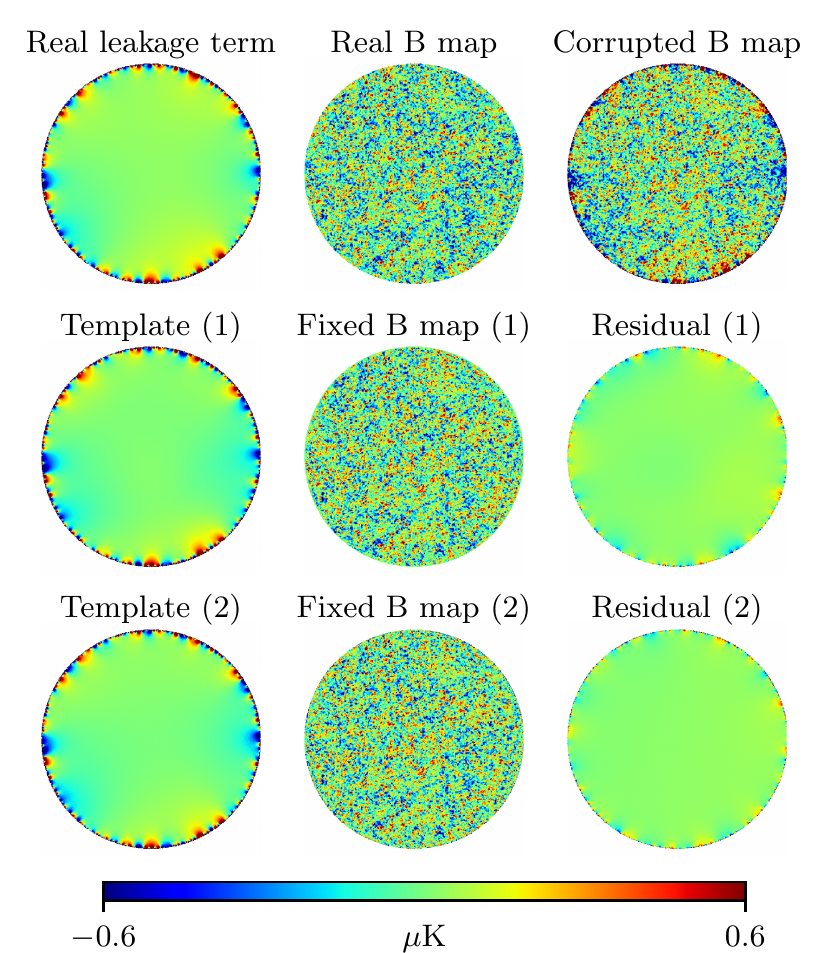}
    \caption{Same as Fig.~\ref{fig:show results} but for a disk mask with
    20$\degree$ radius.}
    \label{fig:show results1}
\end{figure}

We now present examples of correcting the $EB$~leakage on simulated
CMB maps with two different masks, shown in Figs.~\ref{fig:show results}
and~\ref{fig:show results1}.  For this purpose we select a simulated CMB map
with $r=0.05$ from Planck's FFP9 suite.  Both figures show the true signals in
row~1 for reference and the results of correction in rows~2 and~3.

The belt region shown in Fig.~\ref{fig:show results} was defined to be
$20\degree$ in width and $2\degree$ in height. The two methods give similar
leakage templates, and additionally reproduce the real leakage term well. As a
measure of similarity, we compute the cross-correlation between the real
$B$~map and the contaminated one to be only~20$\%$, whereas after correction,
the $B$~map corrected by method~1 gives~86$\%$ correlation with the real
$B$~map, and that of method~2, 66$\%$. While method~1 leads to better
correction on larger scales, method~2 captures the small scale leakage better,
as can be seen in the right panels of rows~2 and~3.

We then repeat this test by instead using a disk-shaped region with
20$\degree$ radius as shown in Fig.~\ref{fig:show results1}. This time, the
cross-correlation between the real $B$~map and the contaminated one is 70$\%$,
whereas after correction, method~1 leads to~97.7$\%$ correlation of the fixed
$B$~map with the real one, and that of method~2 gives~97.6$\%$, in strong
agreement with one another, as well as with the real $B$~map. A glance at the
figures makes clear that most of the interior of the map is significantly
contaminated, which is captured well by the templates.  Given the small
fraction of the edge area in comparison to the whole region, the cross
correlations are only marginally influenced by the edge, especially after
correction. Also note that the cross-correlations are associated with a given
mask, and are not comparable across masks.

Further tests will show that method~2 gives relatively smaller error at the
desired multipole range (as elaborated in Sec.~\ref{sub:post apo}), whereas
method~1 mainly involves the correction of the large-scale features (see also
the smoothness of the template by method 1 in Fig.~\ref{fig:show results}).
Therefore, method~2 will be the default method for the rest of this work.
However, as was the purpose of this section, we point out that method~1 can
perform better in the case of narrow regions, where the edge condition becomes
relatively more important.

\subsection{Advantages of correction in pixel domain}\label{sub:advantage}

Concluding this section, we summarize the advantages a correction of
$EB$~leakage in the pixel domain has over conventional methods that only
recover the $EE$ and $BB$ power spectra.

\begin{itemize}
    \item 
Both methods 1 and 2 operate only in the pixel domain without involving the
power spectrum, i.e., they are independent of assumptions on the $B$-mode
angular power spectrum, and therefore should be considered an
\textit{additional} contribution to existing polarized power spectrum
reconstruction methods.
    \item
As will be shown, another important advantage of pixel domain correction is
that it is very easy to deal with noise, because there noise and CMB are added
linearly, and our proposed correction methods are also linear.
    \item
Since we have already corrected the $EB$~leakage in the pixel domain, the
challenge to arrive at an estimation of the $E$- or $B$-mode power spectrum
simplifies to estimating the angular power spectrum of a \textit{scalar} field
given a mask.  This problem has been intensively studied by many authors,
e.g.,~\cite{ 1997PhRvD..55.5895T, 2001PhRvD..64f3001T, 2002ApJ...567....2H,
2004MNRAS.348..885E, 2004ApJ...609....1J, 2004PhRvD..70h3511W,
2004ApJS..155..227E, 2005JCAP...11..001P, 2006ApJ...645L..89S,
2008StMet...5..289A, 2008PhRvD..77l3539I, 2009PhRvD..79l3515G,
2009MNRAS.400..463G, 2009ApJ...697..258J, 2011AA...536A...5Z,
2012ApJ...750L...9K, 2013AA...550A..15S, 2014MNRAS.440..957M}. This idea is
implemented in Sec.~\ref{sub:with master}, which gives an excellent
reconstruction result.
    \item
As was seen, neither method~1 nor~2 requires any apodization of the mask; they
work simply with a top-hat mask. One is thus free to choose any posterior
apodization scheme to improve the $B$-mode angular power spectrum estimation.
This will be presented in Sec.~\ref{sub:post apo}.
\end{itemize}

\section{Testing the level of residual after correction}\label{sec: evaluation
of the residual}

Even with a perfect $EB$~leakage correction, the $B$-mode spectrum obtained
from the cut sky is still different from the known full-sky spectrum, due to
sampling uncertainty (among others).  To focus on the effectiveness of our
methods, in this section, we perform tests that measure which uncertainties to
expect in $B$-mode power spectra \textit{only} from the contribution of the
$EB$~leakage or its correction.  We hereto use CMB simulations from the FFP9
suite~\citep{2016A&A...594A..12P}, which include the scalar, tensor and
non-Gaussian components, as well as a correctly simulated lensing effect.  As
before, we select those with a tensor-to-scalar ratio $r=0.05$ (except for
Sec.~\ref{sub:eb when zero BB}).  For all the tests we will investigate a
disk-shaped sky region of about $47^\circ$ radius, covering roughly $15\%$ of
the sky.  This choice was made with reference to one of the specifications of
the GreenPol experiment \cite{GreenpolWebsitesUCSB,GreenpolWebsitesNBI}.

First, in Sec.~\ref{sub:eb when zero BB}, we perform a null test on a
zero-$B$-mode simulation. We then move on to investigate simulations with
nonzero $B$-modes and compare our results to those obtained from ``purifying"
the $E$- and $B$-modes in Sec.~\ref{sub:with master}. Lastly, we illustrate
how to further optimize these results by different choices of posterior
apodization in Sec.~\ref{sub:post apo}.

\subsection{Zero initial B-mode}\label{sub:eb when zero BB}

\begin{figure}[!hbp]
    \centering
    \includegraphics{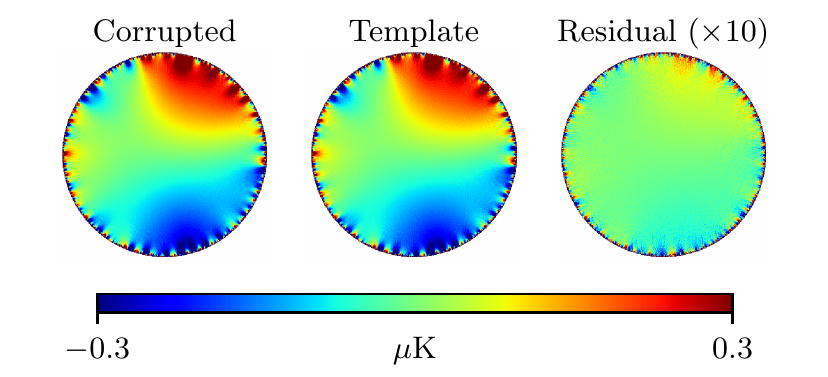}

    \includegraphics{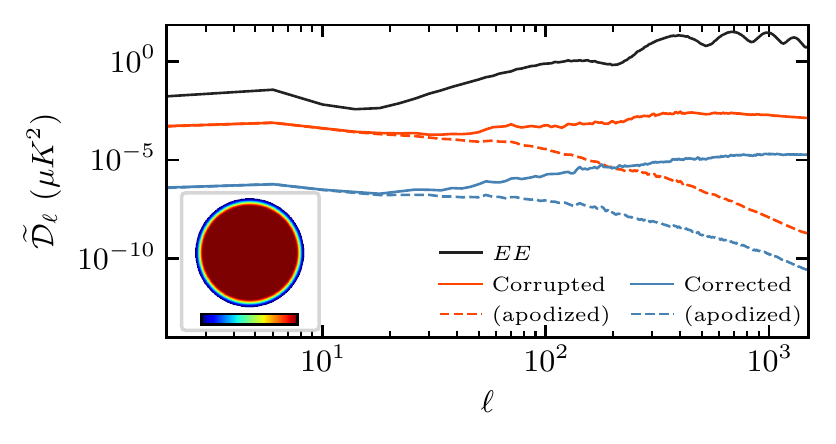}
    \caption{Maps and power spectra after $EB$~leakage correction when the
    input $B$-mode is zero.  \textit{Upper panels}: The corrupted $B$-mode
    (left), the template generated by the recycling method (middle), and the
    residual leakage after correction (multiplied by 10; right). \textit{Lower
    panel}: Comparison of the $EE$ pseudospectrum (black), and the residual
    $BB$ pseudospectra (both binned with $\Delta \ell = 4$) derived from
    either corrupted (red) or corrected (blue) $B$~maps, using either the mask
    (solid) or apodization shown in the inset (dashed). }
    \label{fig:ps results zeroBB}
\end{figure}
We begin with an idealistic test, in which we select a single simulated CMB
map without noise, and manually set the input $B$-mode to zero.  This
automatically marks any detection of a derived $B$-mode signal --- either
before or after correction --- to be due to leakage or residual leakage. After
masking, we attempt to perform an $E/B$ decomposition and subsequently use the
recycling method for the correction of the corrupted $B$~map.  We compare this
to a case where no correction has been done to the corrupted maps. The final
output $B$-mode spectra are then calculated directly from the masked maps, in
two ways:  once where the maps were apodized with a Tukey window for which we
used a taper fraction of~0.1,\footnote{This roughly corresponds to an
apodization length of~$5^\circ$.  In Sec.~\ref{sub:post apo} we study
different window functions for apodization, including the one used here.} and
once where they were not. Those that were apodized were rescaled such that the
spectra are comparable. These spectra correspond to pseudospectra and, as we
mentioned before, are sufficient for highlighting the advantages of our
method, without including sample uncertainties.  As pointed out
by~\cite{Bunn:2002df}, oversampling can help to reduce the leakage due to
pixelization, thus we use $N_{side}=2048$ in this test, and show the results
in Fig.~\ref{fig:ps results zeroBB}. In the upper panels it can be seen that
the leakage from $E$- to $B$-modes is removed almost completely. (We amplify
the residuals by a factor of 10 to make them visible.)  In the bottom panel we
show the angular power spectra of the residual leakages before (red) and after
(blue) correction.  One can see that those whose maps were apodized (dashed
lines) generally give better results than those which were not (solid lines).
The corrected and apodized spectrum gives the best result, which lies up to 12
orders of magnitude below the input $EE$ spectrum.  The other variants are
either worse at large scales (the corrupted $BB$ spectra), or worse at small
scales (without posterior apodization), or both. We already here refer to
Sec.~\ref{sub:post apo}, where we show that our result can be further improved
by about two orders of magnitude by optimizing the apodization.

\subsection{Combination with the MASTER method}\label{sub:with master}

We now extend above test, in which we only considered $BB$ pseudospectra, to
the reconstruction of full $B$-mode spectra. A widely used algorithm to
reconstruct an unbiased full-sky angular power spectrum from the cut sky is
the MASTER method~\citep{2002ApJ...567....2H}.  Our pixel domain $EB$~leakage
correction can be easily combined with the MASTER method (or any other
pseudo-$C_\ell$ method) in the following way.

\begin{figure*}
    \centering
    \includegraphics[width=0.8\textwidth]{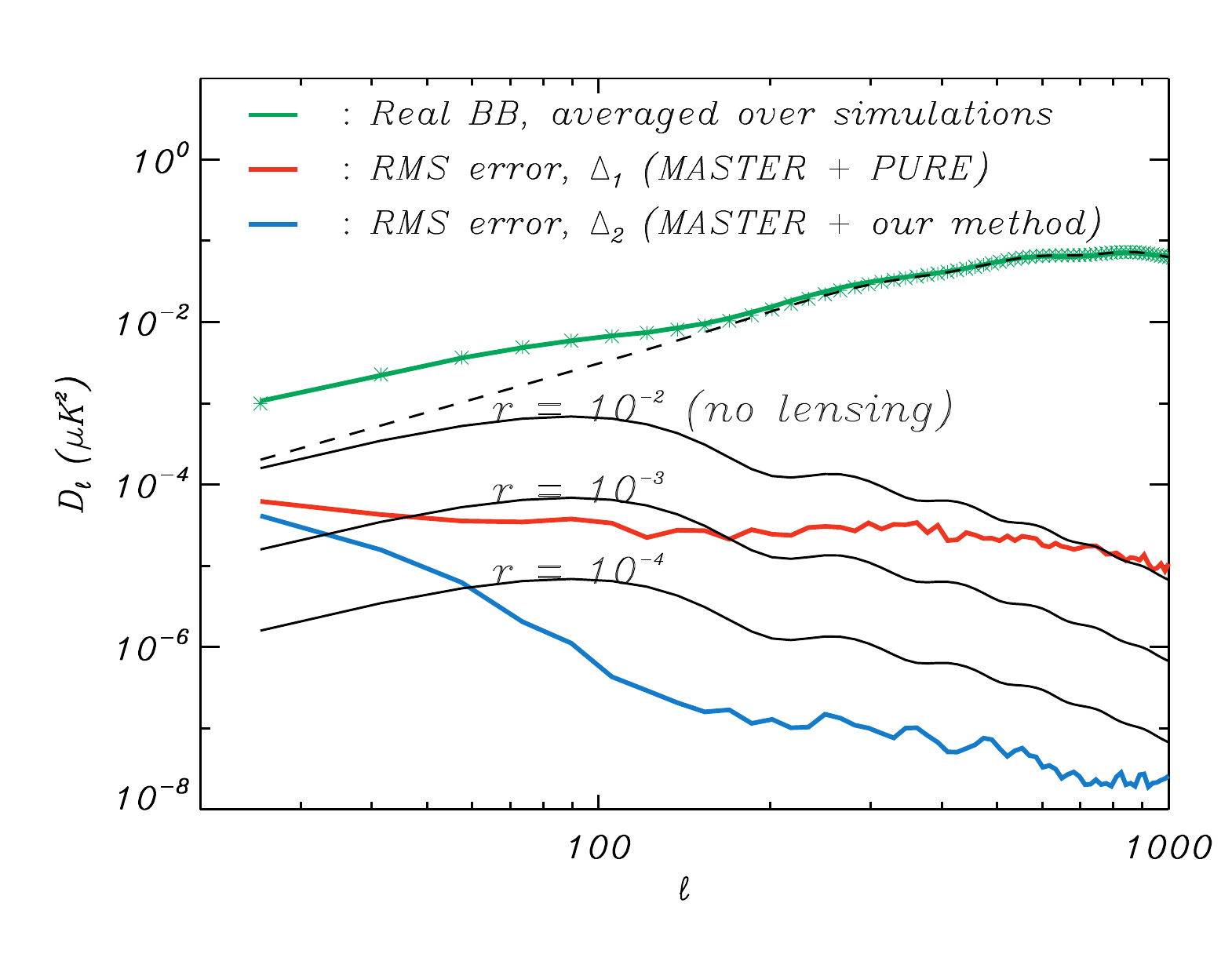}
    \caption{Comparison of the errors of $EB$~leakage correction:
    $\Delta_1(\ell)$ for MASTER+PURE (red), and $\Delta_2(\ell)$ for
    MASTER+our method (blue), see Eqs.~(\ref{equ:master
    test}--\ref{equ:master test2}) and Sec.~\ref{sub:with master} for details.
    Several lines are added for comparison including: the input BB-spectrum of
    $r=0.05$ (green asterisks), the spectrum reconstructed from the real
    B-mode in the available region and averaged over simulations (green
    solid), the expected primordial $B$-mode spectra for
    $r=10^{-2}\sim10^{-4}$ (black solid), and the lensing $B$-mode spectrum
    (black dashed). One can see that our method helps to reduce the error of
    reconstruction by 2--3 orders of magnitudes under the same conditions
    (resolution, simulated maps, sky region, apodization, etc.).}
    \label{fig:with master}
\end{figure*}

\begin{figure}
    \centering
    \includegraphics[width=0.48\textwidth]{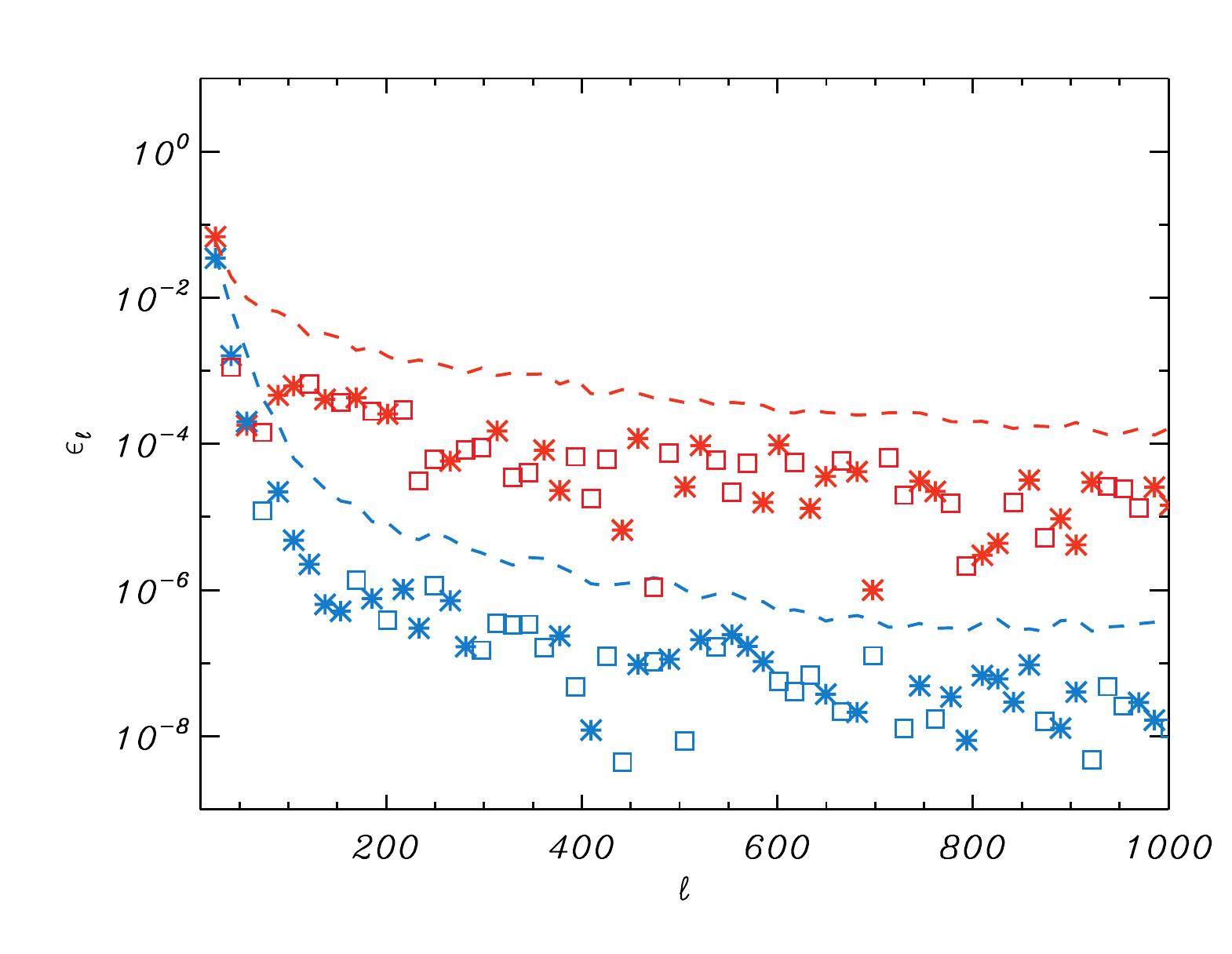}
    \caption{Comparison of the normalized average offsets
    $\epsilon_{1,2}(\ell)$ (see Eq.~(\ref{equ:ave offset})). Red for
    MASTER+PURE and blue for MASTER+our method. The amplitudes of the
    normalized RMS are also shown in dash lines. In both cases, they are
    roughly one order of magnitude higher than the average offsets. The
    simulations are the same ones in Fig.~\ref{fig:with master}. Note that
    this is a logarithmic plot, so we use asterisks and squares to mark
    positive negative values respectively.}
    \label{fig:with master ave}
\end{figure}

We hereto use the Python package, \textsc{pymaster}, of the \textsc{NaMaster}
code~\citep{Alonso:2018jzx, namaster} as an implementation of the MASTER
method to reconstruct the full-sky $BB$ spectrum by two ways for comparison:
one is by using \textsc{NaMaster} with a built-in
purifying~\cite{2006PhRvD..74h3002S} option for the
$B$-mode,\footnote{http://namaster.readthedocs.io/en/latest/sample\_pureb.html.}
whose results are denoted $(C_{\ell}^{BB})_{1,i}$ for 50 different simulations
$i$, and will in the following be referred to as ``MASTER+PURE"; and the
second is to first correct the $B$-mode map by our recycling method, and
subsequently use \textsc{NaMaster} in the \textit{nonpolarized} mode to
reconstruct the full-sky $BB$ spectrum from the corrected $B$~map as
$(C_{\ell}^{BB})_{2,i}$. Lastly, we run MASTER on the \textit{real} $B$~map
for each simulation masked with the same apodization to provide a reference
$(C_{\ell}^{BB})_{0,i}$.  This helps to skip the sampling uncertainty and
focus only on the error of $EB$~leakage correction. The MASTER reconstructions
start from $\ell=16$ and the bin size is also 16, thus the first bin is
centered at $\ell=24$. For each simulation we calculate the differences
between the reconstructions and the reference; subsequently for each we
compute the corresponding RMS and normalized average offsets as:
\begin{align}
\Delta_{1,2}^i(\ell)=&(C_{\ell}^{BB})_{(1,2),i}-(C_{\ell}^{BB})_{0,i};
\label{equ:master test}\\
\Delta_{1,2}(\ell) =&\sqrt{\frac{1}{N_{sim}}
\sum_{i=1}^{N_{sim}}[\Delta_{1,2}^i(\ell)]^2},\label{equ:master test2} \\
\epsilon_{1,2}(\ell) =& \frac{\left\< \Delta_{1,2}^i(\ell) \right\>}
{\left\< (C_{\ell}^{BB})_{0,i} \right\>}\label{equ:ave offset}. 
\end{align}
We plot $\Delta_1(\ell)$ and $\Delta_2(\ell)$ in Fig.~\ref{fig:with master},
and plot $\epsilon_{1,2}(\ell)$ in Fig.~\ref{fig:with master ave} for
comparisons. One can see from Fig.~\ref{fig:with master} that, on average, and
under all the same conditions (resolution, sky region, and apodization), our
method helps to reduce the error of reconstruction by 2--3 orders of magnitude
at higher~$\ell$ with respect to the MASTER+PURE scheme. One can also see from
Fig.~\ref{fig:with master} that the MASTER+PURE method gives uncertainties in
the leakage correction at roughly the level of $r\approx10^{-3}$ for the first
peak and $r\approx10^{-2}$ for higher multipoles, whereas, by an improvement
of 2--3 orders of magnitude, our method ensures that the $EB$~leakage is
suppressed down to a level of $r\approx10^{-4}$--$10^{-5}$ for both the first
peak and higher multipoles.\footnote{We emphasize that this statement holds
\emph{only} for uncertainties arising from $EB$~leakage, and other issues such
as sufficient foreground removal, noise, delensing, sampling uncertainties,
etc., provide additional sources of error.} In Fig~\ref{fig:with master ave},
the normalized average offsets $\epsilon_{1,2}(\ell)$ for MASTER+PURE and
MASTER+our method are compared with each other and with the normalized RMS.
One can see that, in the conditions studied here, our method not only gives
lower RMS, but also gives lower average offsets. Meanwhile, in both cases, the
average offsets are roughly one order of magnitude lower than the
corresponding RMS, so they are both statistically compatible with zero.

Note that in this section we used only $N_{\mathrm{side}}=512$. As stated
before, a higher $N_{\mathrm{side}}$ could help to further improve the
EB-leakage correction~\cite{Bunn:2002df}. Also note that in Fig.~\ref{fig:with
master}, the `C1' apodization from \textsc{NaMaster} with a default
$10\degree$ apodization length was used, without any optimization for either
our method or the pure method. Although this choice is suboptimal, according
to Fig.~17 of~\cite{2009PhRvD..79l3515G}, optimizing the apodization of the
PURE method provides improvements of less than 1 order of magnitude, whereas,
in the section below, we show in Fig.~\ref{fig:windows} that optimizing the
apodization of our method can provide up to 2 orders of magnitude improvement.
Therefore, after optimization is taken into account, we expect our method to
be relatively even better. However, we acknowledge that in realistic
applications, there will be more complicated mask shapes, and no final
conclusion can be drawn about the optimal choice of apodization yet.

\subsection{Optimization of the posterior pixel domain apodization}\label{sub:post apo}

We already know from Figs.~\ref{fig:show results}--\ref{fig:ps results zeroBB}
that the residual $EB$~leakage after correction is most significant at the
edge of the available sky region, and it therefore can be further suppressed
by applying a posterior apodization/window function, where with
\textit{posterior} we mean that the apodization is applied independently of
and after the pixel domain $EB$~leakage correction. Generally speaking, a more
aggressive apodization gives further suppression of the residual leakage, but
at the same time, the  overall signal strength is reduced. In this section we
test different window functions to show how to find a balance between higher
signal and lower residual for the EB-leakage. We use the same mask as in the
two previous subsections as well as Planck FFP9 simulations with $r = 0.05$.

\begin{figure*}
    \includegraphics{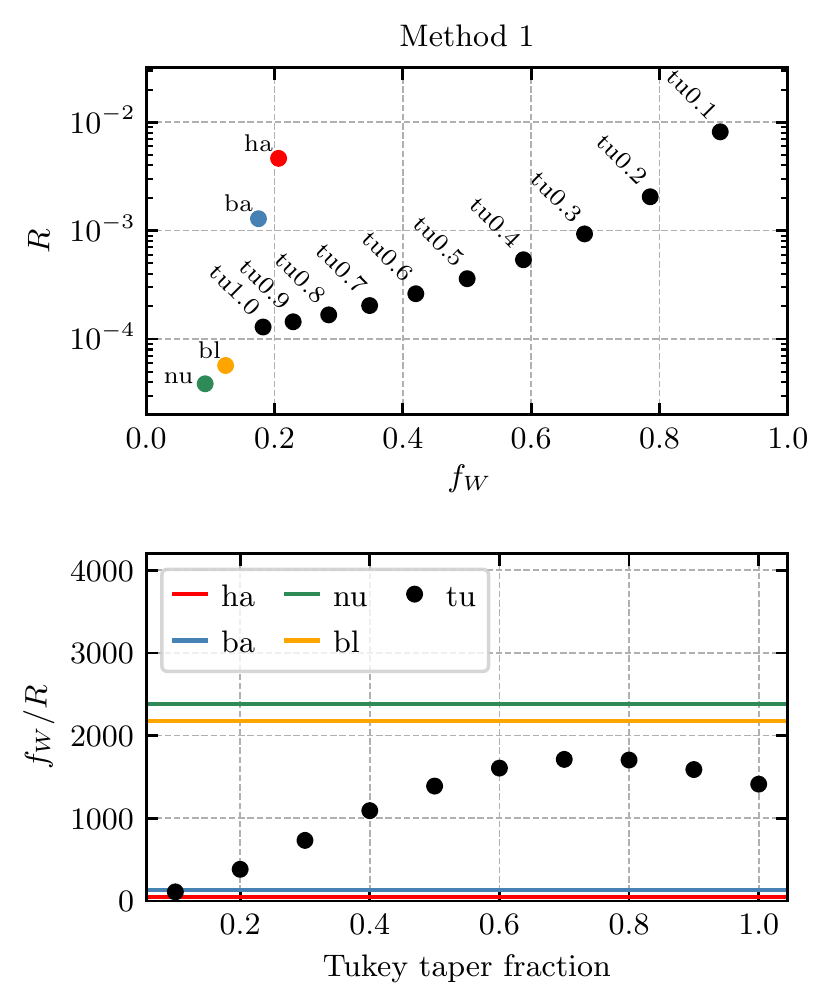}
    \includegraphics{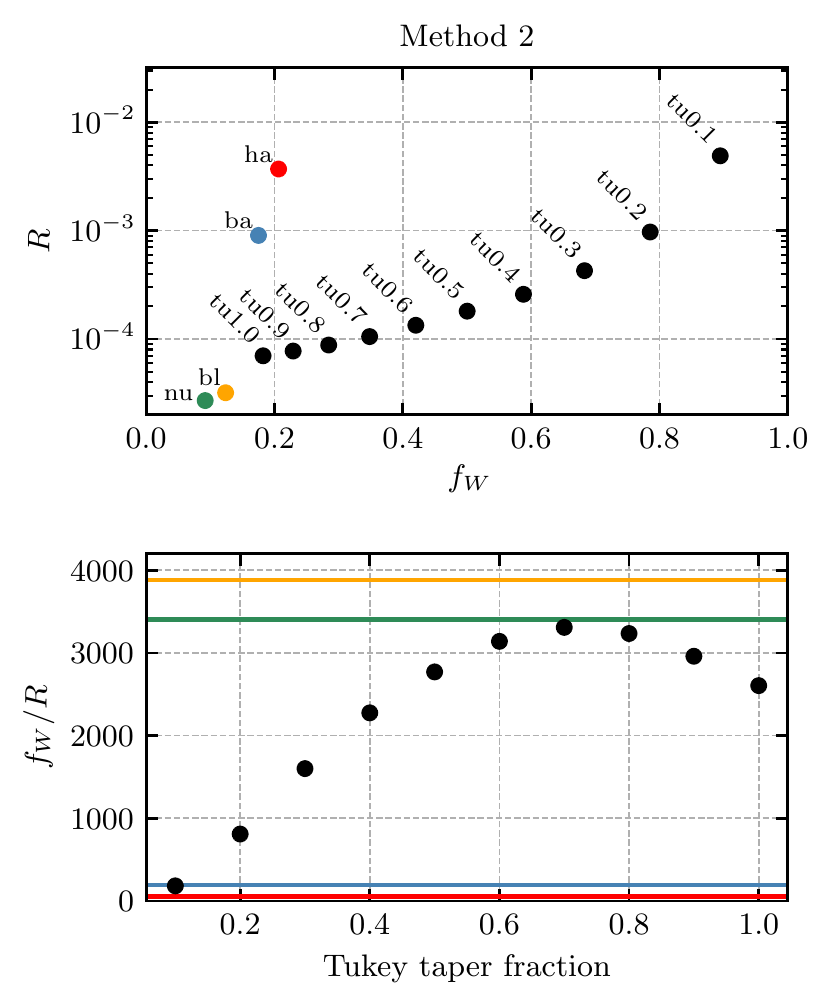}
    \caption{\textit{Upper panels}: comparison of different window functions
    in terms of $f_W$ (Eq.~\ref{equ:fw}) and $R$ (Eq.~\ref{equ:R}).
    The labels are defined in Sec.~\ref{sub:post apo}. Bigger $f_W$ is
    preferred because it keeps more signal power (retaining more information),
    while smaller $R$ is preferred because it means lower residual error of
    correction; note that the recycling method gives roughly $50\%$ lower
    error than method 1. \textit{Lower panels}: ratio $f_W/R$ as a function of
    the taper fraction of Tukey windows, where higher values mean better
    overall performance.}
    \label{fig:windows}
\end{figure*}

Given a symmetric one-dimensional window function defined on the unit
interval, $w(x)$, where $0 \leq x \leq 1$, we construct its corresponding
two-dimensional window function on the available region by $$ W(n) =
w\left(\frac{d(n)}{2d_\mathrm{max}}\right), $$ where $d(n)$ is the distance
from the $n$th pixel to the edge of the mask, and $d_\mathrm{max}$ is the
maximum such distance over all pixels in the available region. Such a
definition ensures that the pixel domain window function is 0 at the edge and
1 at the points that are most distant to the edge. The types of $w(x)$ are
chosen from the following (the abbreviations in brackets are to be used in
Fig.~\ref{fig:windows}):
\begin{itemize}
    \item Hamming (ha) and Tukey windows with taper fractions in increments of
    $0.1$ (tu0.1, etc.) \citep{1455106}
    \item Bartlett window (ba) \citep{533733}
    \item Nuttall window (nu) \citep{1163506}
    \item Exact Blackman window (bl) \citep{6768513,1455106}
\end{itemize} 
Note that the Tukey window is also known as the tapered cosine window; the
conventional cosine window, also known as the Hann window, is recovered with a
taper fraction of $1.0$ (``tu1.0'').

To evaluate the aggressiveness of each window function, we calculate
\begin{align}\label{equ:fw}
f_W = \frac{1}{N}\sum_{n=1}^{N}{W^2(n)}
\end{align}
where $N$ is the total number of pixels in the available region. $f_W$ is an
effective measure of the sky-fraction, normalized such that a top-hat mask
gives $f_W=1$. More aggressive windows remove more power, and we have
\mbox{$f_W\in[0,1]$}.

Given a window/apodization function, the amplitude of the residual leakage
after correction is estimated by $R$, defined as the RMS of the relative
error, averaged over a range of multipoles for all simulations, as follows:

\begin{multline}\label{equ:R}
R = \sqrt{\frac{1}{N_{\mathrm{sim}}\cdot\Delta \ell}\cdot
\left.\sum_{\ell=\ell_1}^{\ell_2}\sum_{i=1}^{N_{\mathrm{sim}}}
\left[\frac{(\widetilde{C}^{BB}_\ell)_i - (\widetilde{C}^{BB,c}_\ell)_i}
{(\widetilde{C}^{BB}_\ell)_i}\right]^2\right.},
\end{multline}
where $(\widetilde{C}^{BB}_\ell)_i$ and $(\widetilde{C}^{BB,c}_\ell)_i$ are
the pseudo powerspectra of the apodized real and corrected $B$~maps for the
$i$-th simulation, and $\Delta\ell=\ell_2-\ell_1+1$. The multipole range used
here is $(\ell_1,\ell_2)=(60,120)$, including the recombination bump
of the $BB$ spectrum.

We evaluate a set of 14 standard window functions, including 10 Tukey windows
with taper fractions in increments of $0.1$. For each window, we plot $R$
vs$.$ $f_W$, and the results for method~1 and method~2 are both shown in
Fig.~\ref{fig:windows}. It is seen that method~2 gives smaller residual error
(lower $R$) than method~1 for each posterior apodization. Furthermore, we also
plot the ratio $f_W / R$ for the different window functions. This ratio is a
simple measure of the overall performance of each window. According to this,
for the mask under investigation, Tukey windows with a taper fraction of
around 0.7, as well as Nuttall and Blackman windows, seem to give the best
EB-leakage correction in both method~1 and~2.

\section{Discussion}\label{sec:disuss}

In this work, we presented two methods (Secs.~\ref{sub:method
1}--\ref{sub:method 2}) that both are capable of correcting the $EB$~leakage
in the pixel domain.  With emphasis on one of them, various tests showed the
effectiveness of these corrections, e.g., the residual error is 2--3 orders of
magnitude lower than an implementation of a conventional method (the
MASTER+PURE scheme). These results are obtained using simple rectangular and
circular masks, and they illustrate the potential improvements possible with
this new method.  The idea of pixel domain $EB$~leakage correction is based on
the idea of $EB$-family decomposition previously proposed
in~\citep{2018JCAP...05..059L, 2018A&A...617A..90L}, which herewith is proved
to be an extremely useful framework for the study of polarization maps.

The advantages of a correction in pixel space are many. In Sec.~\ref{sub:with
master} our $EB$ leakage correction method was combined with MASTER, a
pseudo-$C_l$ method for the reconstruction of a full-sky power spectrum.  We
demonstrated that the results obtained are orders of magnitude better than
without explicit leakage correction.  Our method provides the possibility to
be combined with any pseudo-$C_l$ or maximum likelihood method to improve
their ability for $B$-mode power spectrum reconstruction.

In addition, as shown in Sec.~\ref{sub:post apo}, it is possible to further
reduce the error of power spectrum reconstruction by optimizing the posterior
apodization applied to the corrected $B$ map, see Fig.~\ref{fig:windows}.  We
there explained how to use the large library of one-dimensional window
functions from digital signal processing in CMB science, which provides an
easy way to explore variations two dimensional window functions.

Note that although the rectangular and circular masks studied here are
appropriate for characterizing the general features of the method, they are
simpler than those encountered in practice. More complicated shapes will
result in higher leakage and residuals. Therefore, in the future the residuals
and the bias of the method should be evaluated using simulations with
realistic masks particular to a certain experiment, in order to fully
understand the method's performance on the application in question.

The $EB$~leakage is driven more by large scale structures than by small scale
structures, since small scale structures are locally more confined, and
therefore do not propagate as far. Hence, a satisfactory correction of
$EB$~leakage only requires the $E$-mode to be much larger than the $B$-mode at
large scales, which is always true for the CMB --- also if noise is added,
given that the noise is subdominant compared with the $E$-mode signal at large
scales, which will be the case for upcoming CMB missions.

The methods also enable an easy treatment of noise in power spectrum
reconstructions, because in the pixel domain noise and CMB simply are added
linearly to make up the total signal, and our correction methods are also
linear. Therefore, the $B$-mode residual $\Delta \bm{B}$ after correction is
simply
\begin{align}\label{equ:resi B with noise}
\Delta \bm{B} = \Delta \bm{B}_{\mathrm{CMB}} + \Delta
\bm{B}_{\mathrm{noise}}.
\end{align}
In general, further removal of the noise in the pixel domain is impossible;
however, if one assumes that the noise is Gaussian and uncorrelated with the
CMB, then the two residual terms in Eq.~(\ref{equ:resi B with noise}) are
independent, which means their cross covariance does not contribute to the
overall covariance matrix. With this assumption, one can easily remove the
noise contribution to the angular spectra using one of the standard methods,
by using, e.g., cross spectra~\citep{2003ApJS..148..135H}, noise spectrum
models~\citep{PlanckXV}, null maps obtained from two half-mission
maps~\citep{2016A&A...594A..11P}, or null maps obtained from two
subbands~\citep{2003ApJS..148...63H}. Examples of EB-leakage correction in the
presence of noise can be found in Appendix~\ref{app:with noise}.

To our knowledge, these two methods are the first attempt to provide solutions
to the $EB$~leakage in the pixel domain with negligible computational time
cost. The five main obstacles in the detection of CMB $B$-modes are foreground
removal, delensing, noise, systematics, and the $EB$ leakage.  The present
method to overcome the last also enables the more reliable investigation of
$B$-mode morphology in a local sky region, opening up possibilities to have a
closer look at the remaining obstacles.

\Ack{

We sincerely thank the anonymous referee for the helpful comments. This
research has made use of data product from the
Planck~\citep{Planckdata:online} collaboration, the
\textsc{HEALPix}~\citep{2005ApJ...622..759G} and~\textsc{healpy} packages, and
the \textsc{NaMaster}/\textsc{pymaster}
package~\citep{namaster,Alonso:2018jzx}. This work was partially funded by the
Danish National Research Foundation (DNRF) through establishment of the
Discovery Center and the Villum Fonden through the Deep Space project. Hao Liu
is also supported by the National Natural Science Foundation of China (Grants
No. 11653002, 11653003), the Strategic Priority Research Program of the CAS
(Grant No. XDB23020000) and the Youth Innovation Promotion Association, CAS.

}

\appendix

\section{TWO FORMS OF E/B DECOMPOSITION}

\subsection{The E and B maps}\label{app:eb maps}

Here we briefly review the definition of $E(\bm { n})$ and $B(\bm { n})$ maps.
The Stokes parameters $Q$ and $U$ can be decomposed into spin $\pm2$ spherical
harmonics~\cite{PhysRevD.55.1830, 0004-637X-503-1-1} as follows:
\begin{equation} \label{Q_lm+iU_lm}
    Q(\bm{{n}})\pm i U(\bm{{n}}) = \sum_{l,m}
    a_{\pm2,\ell m}\;{}_{\pm2}Y_{\ell m}(\bm{n}),
\end{equation} where $_{\pm2}Y_{\ell m}(\bm{{n}})$ are the spin $\pm2$
spherical harmonics, and the coefficients $a_{\pm2,\ell m}$ are given by
\begin{equation} \label{a2lm}
    a_{\pm2,\ell m}=\int \left(Q(\bm{{n}})\pm i
    U(\bm{{n}})\right) \,{}_{\pm2}Y^*_{\ell
    m}(\bm{{n}})\,d \bm{{n}}.
\end{equation}
The $E$- and $B$-modes in harmonic space are then formed by
\begin{align}\label{equ:alm-eb}\begin{split}  a_{E,\ell m} &= -(a_{2,\ell m} +
a_{-2,\ell m})/2, \\ a_{B,\ell m} &= i(a_{2,\ell m} - a_{-2,\ell
m})/2,\end{split}
\end{align}
and the pixel domain representations of the $E$- and $B$-modes are
\begin{align}\label{equ:EB-original}\begin{split} 
    E(\bm { n}) &= \sum
    a_{E,\ell m}\,Y_{\ell m}(\bm { n}), \\ B(\bm { n}) &=
    \sum
    a_{B,\ell m}\,Y_{\ell m}(\bm { n}).\end{split}
\end{align}

\subsection{The E- and B-families}\label{app:eb family}

In our work, the $E$- and $B$-families refer to those parts of the Stokes
parameters, $(Q_E,U_E)$ and $(Q_B,U_B)$, that contain only $E$- or $B$-modes
respectively, and satisfy $(Q,U) \equiv (Q_E,U_E) + (Q_B,U_B)$. They are
defined as follows:\footnote{Note that in an early version
of~\cite{2018JCAP...05..059L}, there were misprints in some signs. The
equations here have been corrected.}
\begin{align}
\begin{split}\label{equ:EB-QU_final simple}
\begin{pmatrix}
Q_E \\
U_E
\end{pmatrix}(\bm{n})&= \int
\begin{pmatrix}
G_1  & +G_2 \\
+G_3  & G_4
\end{pmatrix}
(\bm{ n},\bm{ n}')
\begin{pmatrix}
Q \\
U
\end{pmatrix}(\bm{n}')\, d \bm { n}' \\
\begin{pmatrix}
Q_B \\
U_B
\end{pmatrix}(\bm{ n})&= \int
\begin{pmatrix}
G_4  & -G_3 \\
-G_2  & G_1
\end{pmatrix}
(\bm{ n},\bm{ n}')
\begin{pmatrix}
Q \\
U
\end{pmatrix}(\bm{ n}')\,d \bm {n}',
\end{split}
\end{align}
where the $G_{1-4}$ functions are defined as:
\begin{align}
\begin{split}G_{1}(\bm{ n},\bm{ n}') &= \sum_{l,m} F_{+,\ell m}(\bm{
    n})F^*_{+,\ell m}(\bm{ n}'), \\
    \quad G_{2}(\bm{ n},\bm{ n}') &= \sum_{l,m} F_{+,\ell m}(\bm{
    n})F^*_{-,\ell m}(\bm{ n}'), \\
    G_{3}(\bm{ n},\bm{ n}') &= \sum_{l,m} F_{-,\ell m}(\bm{
    n})F^*_{+,\ell m}(\bm{ n}'),\\ \quad G_{4}(\bm{ n},\bm{ n}') &=
    \sum_{l,m} F_{-,\ell m}(\bm{ n})F^*_{-,\ell m}(\bm{ n}'),
    \end{split}
\end{align}
and the $F_{+,-}$ functions are defined in terms of the spin-2 spherical
harmonics as:
\begin{align}
\begin{split}
\label{equ:define F}
    F_{+,\ell m}(\bm{ n}) &= -\frac{1}{2} \left[{}_{2}Y_{\ell m}(\bm{ n}) +
    {}_{-2}Y_{\ell m}(\bm{ n})
    \right], \\
    F_{-,\ell m}(\bm{ n}) &= -\frac{1}{2i}
    \left[{}_{2}Y_{\ell m}(\bm{ n}) - {}_{-2}Y_{\ell m}(\bm{ n}) \right].
\end{split}
\end{align}
Note that $G_i$ are real and $G_2 = G_3$.

The $E$- and $B$-families can also be conveniently calculated by setting
$a_{lm}^B$ or $a_{lm}^E$ to zero, and running a standard inverse
transform using \textsc{HEALPix}. For more details of these two families,
see~\cite{2018JCAP...05..059L, 2018A&A...617A..90L}, also discussed
in~\cite{2018arXiv180711940R}.

\section{DETAILED INTRODUCTION OF METHODS 1 AND 2}\label{app:intro 1 2}

\subsection{Method 1: Diffusive inpainting}\label{appsub:method 1}

In~\cite{2011PhRvD..83h3003B}, it was shown that the ambiguous mode $\psi$
which represents the mixing between the localized $E$ and $B$ estimators
satisfies the spherical bi-Laplacian equation
\begin{align}
    \nabla^2 (\nabla^2 + 2) \psi = 0,
    \label{eq:bilaplace}
\end{align}
subject to homogeneous Neumann and Dirichlet boundary conditions at the edge
of the known region. Assuming that the power in the $E$-mode dominates the
power in the $B$-mode, the purified $B$-mode, calculated by removing the
ambiguous mode $\psi$ from the corrupted $B$-mode, is a good approximation
to the true $B$-mode.

We here simplify the approach by replacing the bi-Laplacian equation by the
Laplacian equation and neglecting the Neumann boundary
conditions.\footnote{The solutions of the simplified Laplacian problem retain
the basic large-scale structure of the bi-Laplacian solutions, however,
small-scale structures can be neglected.} In this case, a simple numerical
solution is the relaxation method. It is important to note that it is possible
to work also with standard full-sky definitions of the $E$- and $B$-mode,
which is more convenient.

It was shown by~\cite{2010A&A...519A.104K} that the $EB$~leakage is most
significant at the edge of the mask, which naturally provides a reliable
boundary condition. Therefore, it is tempting to solve the $EB$~leakage by
relaxation methods using a boundary constraint. This is implemented by
diffusive inpainting~\citep{2014A&A...571A..24P, 2016A&A...594A..17P}, in
which sky pixels\footnote{For this we have used the \textsc{HEALPix} package
(\url{http://healpix.sourceforge.net}) and therefore adopted their
pixelization scheme. However, the method is not tailored to function with that
pixelization only.} are iteratively replaced by the average of their
neighbors, except for the pixels on the boundary. This results in a zero
Laplacian solution that is subject to the given boundary condition.

Therefore, method 1 runs as introduced in section~\ref{sub:method 1}.

\subsection{Method 2: Recycling the E-mode}\label{appsub:method 2}

While the method introduced above provides a means to roughly remove smoothly
distributed leakage, we here suggest a second method which also accounts for
smaller scale structure well inside the unmasked region. Since in this method, the
corrupted component will be reused for correction, it will be referred to as
the ``recycling method''.

Before describing the method we introduce our notation. As discussed
in~\cite{2018JCAP...05..059L, 2018A&A...617A..90L} and briefly reviewed in
Appendix~\ref{app:eb family}, the polarized sky signal can be decomposed into
the $E$- and $B$-families as:
\begin{equation}\label{equ:eb-family}
(Q,U) = (Q_E,U_E)+(Q_B,U_B),
\end{equation} 
where $(Q_E,U_E)$ stems only from the $E$-mode, and $(Q_B,U_B)$ only from the
$B$-mode. This decomposition forms the basis of the recycling method.

\begin{figure}[!b]
    \centering
    \includegraphics[width=0.48\textwidth]{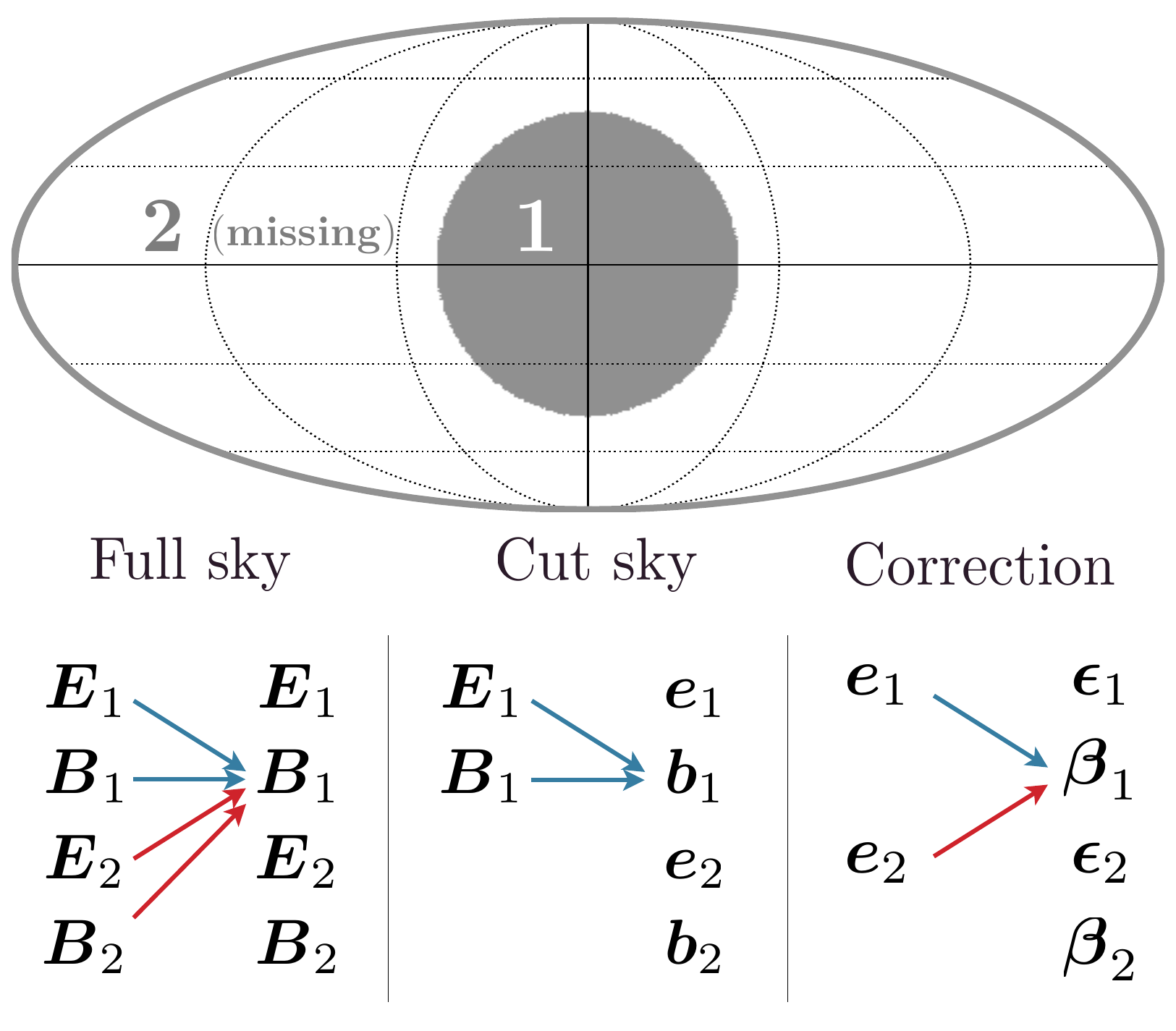}
    \caption{Illustration of recycling method, see Sec.~\ref{sub:method 2} for
    explanations.  Arrows denote contributions from one quantity to another, a
    notation adopted for the equations in the text. }
    \label{fig:principle of method 2}
\end{figure}

Consider a sky map, divided into two regions as shown by
Fig.~\ref{fig:principle of method 2}.
Its polarization signal can be decomposed according to
Eq.~(\ref{equ:eb-family}).  Conventionally the polarization vector is
introduced as $\bm{P}=(Q,U)$, such that $(Q_E,U_E)$ and $(Q_B,U_B)$ can be
denoted as $\bm{P}_E$ and $\bm{P}_B$.  In the following we wish to describe
$\bm{P}_E$ and $\bm{P}_B$ in region 1 and 2 separately, which suggests the
obvious notation $\bm{P}_{E_{1,2}}$ and $\bm{P}_{B_{1,2}}$. However, for
visual simplicity we shorten the notation as follows:
\begin{align}\label{equ:notation}
\begin{split}
&\bm{E}_1\equiv \bm{P}_{E_1}\equiv(Q_E,U_E)_1; \quad \bm{B}_1\equiv\bm{P}_{B_1}
\equiv(Q_B,U_B)_1; \\
&\bm{E}_2\equiv \bm{P}_{E_2}\equiv(Q_E,U_E)_2; \quad \bm{B}_2\equiv\bm{P}_{B_2}
\equiv(Q_B,U_B)_2.
\end{split}
\end{align}
Each Stokes component of $\bm{E}_i$ and $\bm{B}_i$ should be regarded as a
full-sky map whose values at pixels outside region $i$ are zero.  Therefore,
the sum $\bm{E}_1+\bm{B}_1+\bm{E}_2+\bm{B}_2$ again forms the input polarized
sky maps $\bm{P}$. Both $\bm{E}$ and $\bm{B}$ arise from integrating $\bm{P}$
over the full sky.  In this context, each of the quantities in the sum, while
describing only part of the sky, can be thought of as receiving contributions
from all, $\bm{E}_1$, $\bm{B}_1$, $\bm{E}_2$, and $\bm{B}_2$.  The lower-left
panel in Fig.~\ref{fig:principle of method 2} illustrates this process, and in
line with this sketch, we introduce a symbolic notation in which the
contributing terms are denoted with arrows.  We write the corresponding
equation describing the contributions to quantity $\bm{X}$ as
\begin{align}
\label{equ: method2 region1 E}
\bm{X} \equiv (\bm{E}_1 \rightarrow \bm{X}) + (\bm{B}_1 \rightarrow
\bm{X}) + 
(\bm{E}_2 \rightarrow \bm{X}) + (\bm{B}_2 \rightarrow \bm{X}),
\end{align}
where $\bm{X}$ can stand for either of $\bm{E}_1$, $\bm{B}_1$, $\bm{E}_2$ or
$\bm{B}_2$, and in the following we will refer to a bracketed term as a
contributor.  Note that since in practice, region 2 will be the missing part
of the sky (due to either a mask or incomplete observation of the sky),
$\bm{E}_2$ and $\bm{B}_2$ will only be used in the following discussion, but
not in any of the computations presented later.  In this notation we are able
to define a set of rules (Appendix~\ref{app:conservation}) providing detailed
relations between contributors.  In this framework we describe the
$EB$~leakage, and we study relations among the contributors to arrive at a
solution for its correction.

If only region~1 is available, the contributions of $\bm{E}_2$ and
$\bm{B}_2$ obviously disappear, as indicated in the lower middle panel of
Fig.~\ref{fig:principle of method 2}. Consequently, in Eq.~(\ref{equ:
method2 region1 E}) the 3rd and 4th terms disappear, and Eq.~(\ref{equ:
method2 region1 E}) reduces to:
\begin{align}\label{equ:method2 cutsky}
\bm{x} \equiv (\bm{E}_1 \rightarrow \bm{x}) + (\bm{B}_1 \rightarrow
\bm{x}),
\end{align}
where $\bm{x}$ can be either of $\bm{e}_1$, $\bm{b}_1$, $\bm{e}_2$ or
$\bm{b}_2$.  Quantities arising from these incomplete sums are denoted by
lower-case Latin letters, and are what we previously referred to as corrupted.
Obviously, these quantities are generally different from the
corresponding real quantities $\bm{X}$.

Focusing on the corrupted component $\bm{b}_1$, the two contributors that form
it are $(\bm{E}_1 \rightarrow \bm{b}_1)$ and $(\bm{B}_1 \rightarrow
\bm{b}_1)$, as marked by the blue arrows in the lower, middle panel of
Fig.~\ref{fig:principle of method 2}. These two contributors have distinct
meanings: $(\bm{B}_1 \rightarrow \bm{b}_1)$ contains the $B$-to-$B$
deformation, which can be corrected in the angular power spectrum, e.g., by
the MASTER method~\citep{2002ApJ...567....2H}; the contributor $(\bm{E}_1
\rightarrow \bm{b}_1)$ is the $EB$~leakage. It is this term which we attempt
to correct for in this work. However, since $\bm{E}_1$ is unknown in the case
of partial sky coverage, the true leakage term $(\bm{E}_1
\rightarrow \bm{b}_1)$ is generally not available. Nevertheless,
with some approximations, we shall show how to remove this leakage in the
pixel domain to a highly sufficient degree.

We hereto restrict ourselves to the $E$-family output of the cut-sky case,
$\bm{e}_1$ and $\bm{e}_2$, which together form a full-sky map of the
$E$-family.  We decompose this map again in terms of $E$- and $B$-families, as
shown in the lower-right panel of Fig.~\ref{fig:principle of method 2}.
However, this will not produce any $B$-family output (except for numerical and
pixelization errors), which in terms of the contributors is written as
\begin{align}
\label{equ:betasumzero}
\bm{\beta}_1 = (\bm{e}_1\rightarrow \bm{\beta}_1) &+ (\bm{e}_2\rightarrow
\bm{\beta}_1)=0; \\  -(\bm{e}_1 \rightarrow \bm{\beta}_1)
&=(\bm{e}_2\rightarrow \bm{\beta}_1)\ne 0.
\end{align}
Here and in the figure, lower-case Greek letters denote the $E$-
and $B$-families from the corrupted maps $\bm{e}_1 + \bm{e}_2$.

In the CMB, $E$-modes clearly dominate over $B$-modes.  The
observation\footnote{For the two cases discussed in Sec.~\ref{sub:examples} we
find correlation coefficients of $0.99$ and $0.97$, respectively.} that then
$\bm{E}_1 \approx \bm{e}_1$ enables us to reason that the morphology of the
$EB$~leakage term $(\bm{E}_1 \rightarrow \bm{b}_1)$ is well approximated by
the contributor $(\bm{e}_1 \rightarrow\bm{\beta}_1)$.  In fact, we expect an
approximate proportionality between the two contributors (see
Appendix~\ref{app:more discu}) such that the $E$-to-$B$~leakage can be
corrected in the pixel domain by linearly removing the contributor $(\bm{e}_1
\rightarrow\bm{\beta}_1)$ from $\bm{b}_1$.  In short, we recycle a product of
one corrupted component, $\bm{e}_1$, for the correction of another, $\bm{b}_1$.

Therefore, method 2 runs as introduced in section~\ref{sub:method 2}.

\section{The symbolic system behind the recycling method}\label{app:conservation}

The description of the recycling method presented in Appendix
\ref{appsub:method 2} is based on a purely symbolic representation. In this
formulation, we here provide a complete set of relations among the quantities.
For convenience and consistency, we continue to use the notation from
Eq.~(\ref{equ:notation}).

Firstly we have the total conservation rules, which were given in
Eq.~(\ref{equ: method2 region1 E}) and rewritten below:
\begin{align}\label{equ:total con retype}
\bm{X} \equiv &(\bm{E}_1 \rightarrow \bm{X}) + (\bm{B}_1 \rightarrow
\bm{X}) + 
(\bm{E}_2 \rightarrow \bm{X}) + (\bm{B}_2 \rightarrow \bm{X}),
\end{align}
where $\bm{X}$ is one of $\bm{E}_1$, $\bm{B}_1$, $\bm{E}_2$ or $\bm{B}_2$.

As presented in Eq.~(\ref{equ:betasumzero}), the contributors presented
in Sec~\ref{sub:method 2} satisfy the orthogonality rules, which can be
written as:
\begin{align}\label{equ:isolation laws}
(\bm{X}_i\rightarrow\bm{Y}_i) +
(\bm{X}_j\rightarrow\bm{Y}_i) &= 0,
\end{align}
where $\bm{X}$ and $\bm{Y}$ are either $\bm{E}$ or $\bm{B}$, but not the same;
and $i$, $j$ are either 1 or 2 but not the same. For convenience, we extend
Eq.~(\ref{equ:isolation laws}) as follows:
\begin{align}
\begin{split}
\label{equ:isolation laws long}
(\bm{E}_1\rightarrow\bm{B}_1) +
(\bm{E}_2\rightarrow\bm{B}_1) &= 0, \\
(\bm{E}_1\rightarrow\bm{B}_2) +
(\bm{E}_2\rightarrow\bm{B}_2) &= 0, \\
(\bm{B}_1\rightarrow\bm{E}_1) +
(\bm{B}_2\rightarrow\bm{E}_1) &= 0, \\
(\bm{B}_1\rightarrow\bm{E}_2) +
(\bm{B}_2\rightarrow\bm{E}_2) &= 0.
\end{split}
\end{align}
These rules follow from the orthogonality between the $E$- and $B$-families.

Combining the total conservation rules with the orthogonality rules, one gets
the inner conservation rules as follows:
\begin{align}\label{equ:inner conservation laws}
(\bm{X}_i\rightarrow\bm{X}_i) +
(\bm{X}_j\rightarrow\bm{X}_i) &=
\bm{X}_i,
\end{align}
where $\bm{X}$ is either $\bm{E}$ or $\bm{B}$, and $i$, $j$ are either 1 or 2
but not the same. Again we extend this equation for convenience as:
\begin{align}
\begin{split}
\label{equ:inner conservation laws long}
(\bm{E}_1\rightarrow\bm{E}_1) +
(\bm{E}_2\rightarrow\bm{E}_1) &=
\bm{E}_1,\\ 
(\bm{E}_1\rightarrow\bm{E}_2) +
(\bm{E}_2\rightarrow\bm{E}_2) &=
\bm{E}_2,\\ 
(\bm{B}_1\rightarrow\bm{B}_1) +
(\bm{B}_2\rightarrow\bm{B}_1) &=
\bm{B}_1,\\ 
(\bm{B}_1\rightarrow\bm{B}_2) +
(\bm{B}_2\rightarrow\bm{B}_2) &=
\bm{B}_2.
\end{split}
\end{align}

When one connects the full-sky quantities ($\bm{E}$ or $\bm{B}$) with cut-sky
quantities ($\bm{e}$ or $\bm{b}$), one has the completeness rules as follows:
\begin{align}\label{equ:extra conservation laws2}
\bm{E}_1+\bm{B}_1 &=
\bm{e}_1+\bm{b}_1,\\ \nonumber
\bm{E}_2+\bm{B}_2 &=
\bm{e}_2+\bm{b}_2,
\end{align}
which follow from the completeness of the spin-2 spherical harmonics.
In particular, when region 2 is unavailable, we have
\begin{align}\label{equ:extra conservation laws21}
\bm{E}_2+\bm{B}_2 &=
\bm{e}_2+\bm{b}_2=0.
\end{align}

All $E$-to-$E$, $B$-to-$B$, $E$-to-$B$, and $B$-to-$E$ leakages can be
formally described by, and are also subject to, the symbolic system
represented by Eqs.~(\ref{equ:total con retype}--\ref{equ:extra
conservation laws21}). Note that the symbolic system in this section does not
contain any approximation. The approximation needed for the recycling method
is contained in Eq.~(\ref{equ:D6}) in Appendix~\ref{app:more discu}.\\

Alternatively, all rules presented here can also be expressed in terms of the
equations in Appendix~\ref{app:eb family}. With the definitions:
\begin{align}
G_E\equiv
\begin{pmatrix}
G_1  & +G_2 \\
+G_3  & G_4
\end{pmatrix} ;~
G_B\equiv
\begin{pmatrix}
G_4  & -G_3 \\
-G_2  & G_1
\end{pmatrix},
\end{align}
we can express the contributors as, e.g.,
\begin{align}
\bm{X}=&\int G_{\bm{X}}(\bm{n},\bm{n}') \bm{P}(\bm{n}')\, d\bm{n}'; \\
\bm{x}=&\int_1 G_{\bm{X}}(\bm{n},\bm{n}') \bm{P}(\bm{n}')\, d\bm{n}'; \\
(\bm{X}_i\rightarrow\bm{Y}_j) =& \left[\int_i G_{\bm{Y}}(\bm{n},\bm{n}')
\bm{X}(\bm{n}')\,d\bm{n}' \right]_j\;,
\end{align}
where $\int_i$ denotes the integration over region~$i$ only, and $[...]_j$
denotes the restriction of the evaluated quantity in region~$j$.

\section{About the linear fitting for recycling method}\label{app:more discu}

In the recycling method, we use linear fitting to determine the factor that
connects the template $(\bm{e}_1\rightarrow\bm{\beta}_1)$ to the real leakage
$(\bm{E}_1\rightarrow\bm{b}_1)$. Here we provide more details on why this can
be done by linear fitting.

For a cut sky map, only region 1 is available, thus we have
\begin{align}
\bm{E}_1 + \bm{B}_1 = \bm{e}_1 +
\bm{b}_1.
\end{align} 
Assuming there is no initial $B$-mode (like the assumption in Sec.~\ref{sub:eb
when zero BB}), or $B\ll E$ and $B$ therefore can be neglected, then we have
\begin{align}\label{equ:identical maps}\begin{split}
\bm{E}_1=\bm{e}_1+\bm{b}_1.\end{split}
\end{align}
Since now there is only one input component $\bm{E}_1$, we automatically get
\begin{align}
\label{equ:D3a}
\bm{e}_1&=(\bm{E}_1\rightarrow\bm{e}_1),
\\
\label{equ:D3b}
\bm{b}_1&=(\bm{E}_1\rightarrow\bm{b}_1).
\end{align}
Since there is no signal in region 2, we also have
(cf.~Eq.~(\ref{equ:extra conservation laws21}))
\begin{align}
\label{equ:D4}
\bm{e}_2+\bm{b}_2=0;
\end{align}
however, note that at the same time we have $\bm{e}_2\ne0$ and
\mbox{$\bm{b}_2\ne0$}.

Eqs.~(\ref{equ:D3b}--\ref{equ:D4}) tell us that
\begin{align}\label{equ:D5}
\begin{split}
\bm{b}_1 \equiv&
(\bm{e}_1\rightarrow\bm{b}_1) +
(\bm{b}_1\rightarrow\bm{b}_1) + 
(\bm{e}_2\rightarrow\bm{b}_1) +
(\bm{b}_2\rightarrow\bm{b}_1) \\  
=&
(\bm{e}_1\rightarrow\bm{b}_1) +
(\bm{b}_1\rightarrow\bm{b}_1) = (\bm{E}_1\rightarrow\bm{b}_1).
\end{split}
\end{align}
Since the contributor $(\bm{b}_1\rightarrow\bm{b}_1)$ represents the
$B$-to-$B$ leakage, we can expect it to have similar morphology to the input
$B$-mode even with a mask. Thus we have
\begin{align}\label{equ:D6}
(\bm{b}_1\rightarrow\bm{b}_1) \approxprop \bm{b}_1,
\end{align}
where $\approxprop$ stands for approximate proportionality. Thus according to
Eq.~(\ref{equ:D5}), as long as the amplitudes of $\bm{b}_1$ and
$(\bm{b}_1\rightarrow\bm{b}_1)$ are not close to each other (which is
observed), we find
\begin{align}\label{equ:D7}
(\bm{e}_1\rightarrow\bm{b}_1) \approxprop
(\bm{E}_1\rightarrow\bm{b}_1).
\end{align}
Finally, two contributors are always equal when the input components,
destination regions and output types are all equal, thus we have
\begin{align}\label{equ:D8}
(\bm{e}_1\rightarrow\bm{\beta}_1) = (\bm{e}_1\rightarrow\bm{b}_1)
\approxprop (\bm{E}_1\rightarrow\bm{b}_1).
\end{align}
Since $(\bm{e}_1\rightarrow\bm{\beta}_1)$ is our template, and
$(\bm{E}_1\rightarrow\bm{b}_1)$ is the real $EB$~leakage,
Eq.~(\ref{equ:D8}) says that one can use linear fitting to connect the
template to the real leakage in the recycling method.

It is also easy to explain why we choose $(\bm{e}_1 \rightarrow
\bm{\beta}_1)$ as the template, but not $(\bm{b}_1
\rightarrow \bm{\beta}_1)$: when the input $B$-mode is not zero, $(\bm{b}_1
\rightarrow \bm{\beta}_1)$ will contain the real input $B$-mode, but
$(\bm{e}_1 \rightarrow \bm{\beta}_1)$ will not.

\section{The EB-leakage with noise}\label{app:with noise}

Here we reproduce Fig.~\ref{fig:with master}--\ref{fig:with master ave} by
adding white noise with 1 and 10 $\mu$K$\cdot$arcmin amplitudes to each
simulation, respectively. We run the test in a completely blind way, so each
program, either MASTER+PURE or MASTER+our method, is unaware of the presence
of noise. Thus the resulting angular power spectrum will automatically include
the noise contribution, and become higher than the primordial one. It is
important to notice that, according to Eqs.~\ref{equ:master
test}--\ref{equ:master test2}, in a blind test, the error of reconstruction
will automatically use the noisy map as reference, i.e., the noise itself is
regarded as a natural part of the input map, and will \emph{not} be removed
here. The advantage of this approach is that it helps us to focus on the
EB-leakage itself.

The results of the tests with noise are presented in Fig.~\ref{fig:with
master1}, where one can see that the green line becomes higher with higher
noise, but the error of EB-leakage correction for both methods is still
reasonably small, so they both give fairly stable EB-leakage correction
with/without noise. We also notice that, when noise increases, the error of
EB-leakage correction increases more slowly for our method than for
MASTER+PURE.
\begin{figure*}
    \centering
    \includegraphics[width=0.48\textwidth]{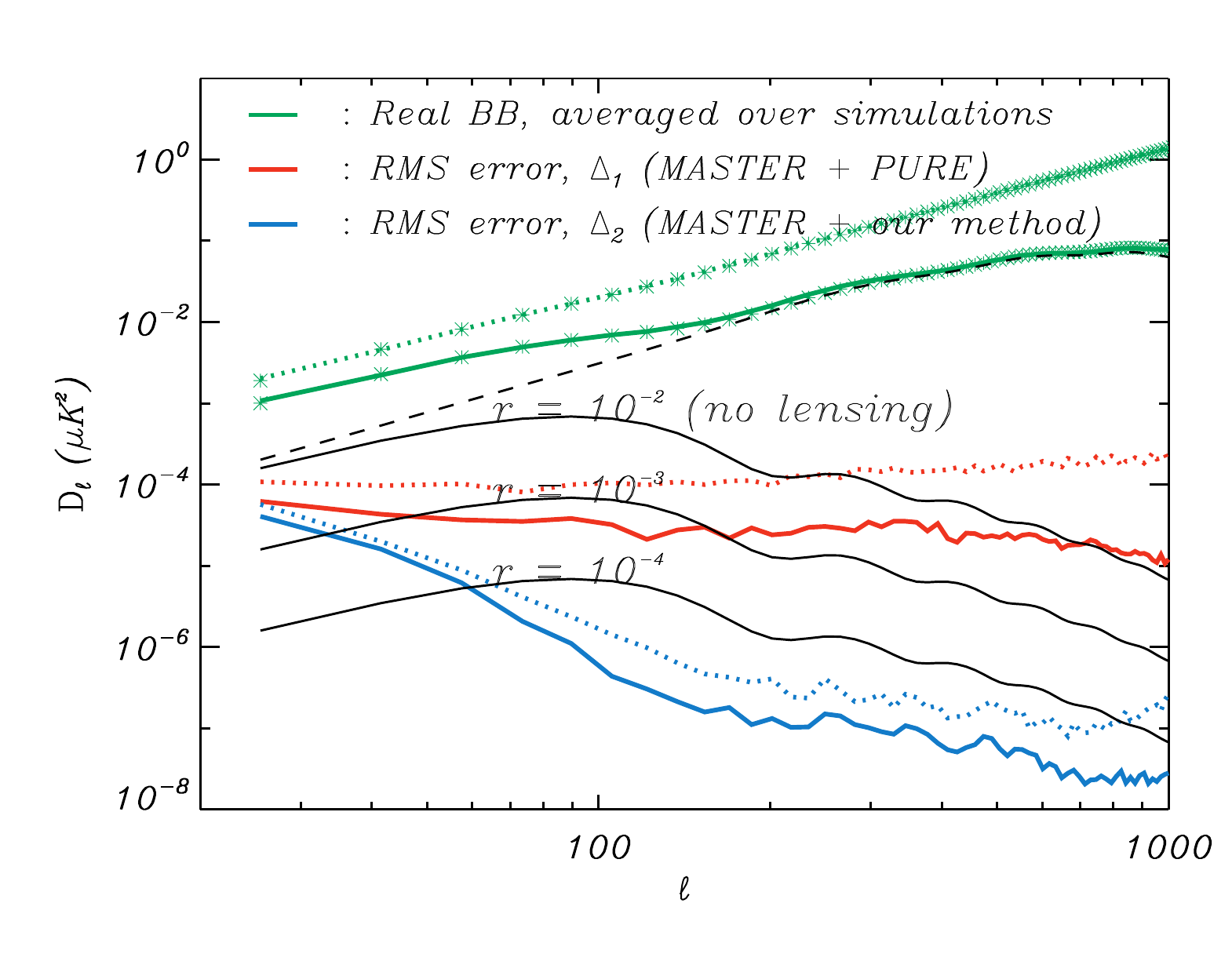}
    \includegraphics[width=0.48\textwidth]{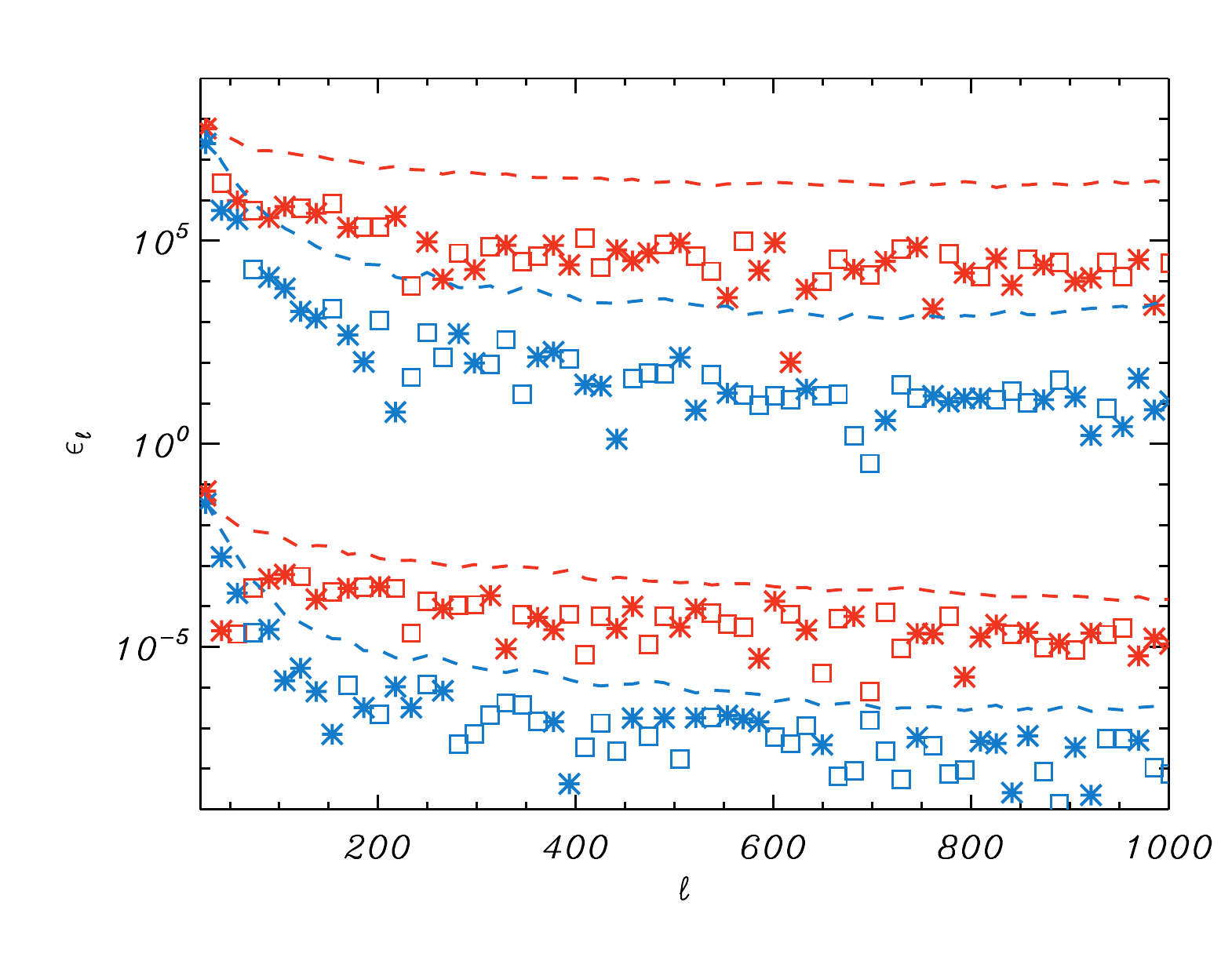}
    \caption{\emph{Left}: Same as Fig.~\ref{fig:with master}, only that white
    noise with 1 and 10 $\mu$K$\cdot$arcmin amplitudes (solid/dotted) is added
    to each simulation respectively. Note that the test is run in a blind way,
    so noise is regarded as natural part of the input maps. \emph{Right}: The
    corresponding $\epsilon_{1,2}(\ell)$, similar to Fig.~\ref{fig:with master
    ave}. The 10 $\mu$K$\cdot$arcmin noise results are up-shifted to make the
    lines visible.}
    \label{fig:with master1}
\end{figure*}


\end{document}